\documentclass[PRL,preprintnumbers,floatfix,aps,notitlepage,nofootinbib,twocolumn,showpacs,amssymb]{revtex4-1}
\usepackage{amsmath,amsfonts,bm}
\usepackage{graphicx}
\usepackage{cancel}
\usepackage[colorlinks, linkcolor={red},citecolor={blue}]{hyperref}
\usepackage{xcolor}

\newcommand{\rs}{r_{\rm s}}

\begin{document}
\title{From black hole mimickers to black holes}
\author{Roberto Casadio}
\email{casadio@bo.infn.it}
\affiliation{Dipartimento di Fisica e Astronomia,
Alma Mater Universit\`a di Bologna,
40126 Bologna, Italy
\\
Istituto Nazionale di Fisica Nucleare, I.S.~FLaG
Sezione di Bologna, 40127 Bologna, Italy}
\author{Alexander Kamenshchik}
\email{kamenshchik@bo.infn.it}
\affiliation{Dipartimento di Fisica e Astronomia,
Alma Mater Universit\`a di Bologna,
40126 Bologna, Italy
\\
Istituto Nazionale di Fisica Nucleare, I.S.~FLaG
Sezione di Bologna, 40127 Bologna, Italy}
\author{Jorge Ovalle}
\email[]{Corresponding author: jorge.ovalle@physics.slu.cz}
\affiliation{Research Centre for Theoretical Physics and Astrophysics,
Institute of Physics, Silesian University in Opava, CZ-746 01 Opava,
Czech Republic.}
\begin{abstract}
We present a simple analytical model for studying the collapse
of an ultracompact stellar object (regular black hole mimicker with infinite redshift
surface) to form a (integrable) black hole, in the framework of General Relativity.
Both initial and final configurations have the same ADM mass, so that the transition 
represents an internal redistribution of matter without emission of energy.
The model, despite being quite idealized, can be viewed
as a good starting point to investigate near-horizon quantum physics.
\end{abstract} 
\maketitle
%
%
%
\section{Introduction}
Ultracompact stellar objects are matter distributions with radii ever so slightly larger
than their gravitational (Schwarzschild) radius.
Therefore their luminosity is subjected to a very large (perhaps even infinite)
redshift and they turn out to be excellent candidates for black hole (BH)
mimickers~\cite{Mazur:2004fk,Mazur:2001fv,Visser:2003ge,Lobo:2005uf}.
Understanding the formation and possible existence of such astrophysical
systems is a pending task.
In fact, there are still many open questions, regarding their stability in particular, 
and what would prevent them from collapsing further into a BH.
\par
One way to avoid this fate is the anti-gravitational effect generated by 
anisotropic pressures within the stellar structure [see Eq.~\eqref{con111} below].
On the other hand, if such ultracompact stellar objects do exist and begin to gain mass
from the surrounding environment, they could eventually become unstable
and collapse to an even more compact and more stable configuration.
In fact, the final stage of the gravitational collapse in General Relativity is quite
generically predicted to be a BH singularity hidden behind the event horizon.
\par
For the above reasons, studying the transition from mimickers to BH
appears an attractive issue.
However, dynamical processes of this type are highly complex due to the
nonlinearity of General Relativity, to the point that numerical calculations often
remain the only viable option.
Nonetheless, in this work, we will describe a simple analytical model
for an ultracompact object (with infinite redshift surface) that further collapses
into an integrable BH, that is a BH characterised by an energy density which
is regular enough to make the mass function vanish at the
centre~\cite{Lukash:2013ts}.
This condition may be sufficient to avoid the existence of
inner horizons~\cite{,Casadio:2021eio,Casadio:2022ndh,Casadio:2023iqt}.
and still preserve some desirable features of regular black
holes~\cite{Carballo-Rubio:2023mvr}.
\par
In the next Section, we briefly review properties of Kerr-Schild spacetimes
which will serve to construct the interior and exterior geometries for a class
of integrable BHs and mimickers in Section~\ref{sec3};
a model for the transition from such mimickers to BHs is then
described in Section~\ref{sec4} and finale remarks are given in
Section~\ref{conc}. 
\section{Kerr-Schild spacetimes}
\label{sec2}
For spherically symmetric and static spacetimes, the general solution of the Einstein field
equations,~\footnote{We use units with $\kappa=8\,\pi\,G_{\rm N}$ and $c=1$.}
\begin{equation}
\label{efe}
R_{\mu\nu}-\frac{1}{2}\, R\,  g_{\mu\nu}
=
\kappa\,{T}_{\mu\nu}\ ,
\end{equation}
can be
written as~\cite{Visser:1995cc}
\begin{equation}
ds^{2}
=
-e^{\Phi(r)}\,f(r)\,dt^{2}+\frac{dr^2}{f(r)}+r^2\,d\Omega^2
\ ,
\label{metric}
\end{equation}
where 
\begin{equation}
f
=
1-\frac{2\,m(r)}{r}
\ ,
\end{equation}
with $m$ the Misner-Sharp-Hernandez mass~\cite{Misner:1964je,Hernandez:1966zia}.
The case $\Phi=0$ corresponds to spacetimes of the so-called
Kerr-Schild class~\cite{kerrchild}, for which Eq.~\eqref{efe} yields the energy-momentum tensor
\begin{eqnarray}
\label{emt}
T^\mu_{\ \nu}
=
{\rm diag}\left[-\epsilon,\,p_r,\,p_\theta,\,p_\theta\right]
\ ,
\end{eqnarray}
with energy density $\epsilon$, radial pressure $p_r$ and transverse pressure
$p_\theta$ given by
\begin{eqnarray}
\label{sources}
\epsilon
=
\frac{2\,{m}'}{\kappa\,r^2}
\ ,
\quad
p_r
=
-\frac{2\,{m}'}{\kappa\,r^2}
=
-\epsilon
\ ,
\quad
p_\theta
=
-\frac{{m}''}{\kappa\,r}
\ ,
\end{eqnarray}
where primes denote derivatives with respect to $r$.
This source must be covariantly conserved, $\nabla_\mu\,{T}^{\mu\nu}=0$,
which yields
\begin{eqnarray}
\label{con111}
p_r'
&=&
-
\left[\frac{{\Phi}'}{2}+\frac{m-r\,m'}{r\,(r-2\,m)}\right]
\left(\epsilon+p_r\right)
+
\frac{2}{r}
\left(p_\theta-p_r\right)
\nonumber
\\
&=&
\frac{2}{r}\left(p_\theta-p_r\right)
\ ,
\end{eqnarray}
where we used $\Phi=0$ and the second of Eqs.~\eqref{sources}.
\par
We also note that the Einstein field equations~\eqref{sources} are linear in the mass function.
Two solutions with $m=m_1(r)$ and $m={m}_2(r)$ can therefore be combined to generate
a new solution with
\begin{eqnarray}
\label{gdks}
m(r)=m_1(r)+m_2(r)
\ . 
\end{eqnarray}
Eq.~\eqref{gdks} represents a trivial case of the so-called gravitational
decoupling~\cite{Ovalle:2017fgl,Ovalle:2019qyi}. 
\par
If we use a metric of the form in Eq.~\eqref{metric} with $\Phi=0$ to describe the interior
and exterior of a stellar object of radius $\rs$, the two regions will join smoothly at $r=\rs$
provided the interior mass function $m$ satisfies
\begin{equation}
\label{c1}
m(\rs)=\tilde{m}(\rs)
\quad
{\rm and}
\quad
m'(\rs)=\tilde{m}'(\rs)\ ,
\end{equation}
where $\tilde{m}$ stands for the exterior mass function and $F(\rs)\equiv\,F(r)\big\rvert_{r=\rs}$
for any function $F(r)$.
Therefore, from Eqs.~\eqref{sources} and~\eqref{c1}, we conclude that the density and radial
pressure are continuous at the boundary $r=\rs$, that is
\begin{equation}
\label{c2a}
\epsilon(\rs)=\tilde{\epsilon}(\rs)
\quad
{\rm and}
\quad
p_r(\rs)=\tilde{p}_r(\rs)
\ ,
\end{equation}
where $\tilde{\epsilon}$ and $\tilde{p}_r$ are the energy density and radial pressure
for the exterior region, respectively.
Finally, notice that the transverse pressure $p_\theta$ is in general discontinuous
across $\rs$.
\section{Black holes and mimickers}
\label{sec3}
The existence of BHs with a single horizon was recently investigated in 
Ref.~\cite{Ovalle:2023vvu}, by combining different Kerr-Schild configurations
like in Eq.~\eqref{gdks}.
We shall here consider some solutions found therein as candidates of
BHs and their mimickers.
\subsection{Interiors}
\label{sec:interior}
In Ref.~\cite{Ovalle:2023vvu}, a non-singular line element representing
the interior of a ultracompact configuration of radius $\rs$ was found
that does not contain Cauchy horizons.
This is given by the metric~\eqref{metric} with $\Phi=0$ and
\begin{equation}
f=
f^\pm(r)
=
1-\frac{2\,m^\pm(r)}{r}
\ ,
\end{equation}
with mass function 
\begin{equation}
\label{m}
m^\pm(r)
=
\frac{r}{2}\left\{
1\pm\left[1-\left(\frac{r}{\rs}\right)^n\right]^{k}
\right\}
\ ,
\quad
0\le r\leq\rs
\ ,
\end{equation}
where $k$ and $n$ are constants.
The analysis of the causal structure of this metric shows that it represents
a BH for $m=m^+(r)$ and an ultracompact configuration with an infinite redshift 
surface for $m=m^-(r)$.
For both cases, the mass $M$ of the system~\footnote{This is not the
ADM mass~\cite{Arnowitt:1959ah}, as we will see in Section~\ref{sec:exterior}.} 
\begin{equation}
\label{Mmim}
M
\equiv
m(\rs)
=
{\rs}/{2}
\ ,
\end{equation}
is contained inside the Schwarzschild radius $\rs=2\,M$.
\par
The energy-momentum $T^{\mu\nu}$ of the source generating
these metrics is simply obtained by replacing the mass function $m^\pm$
in Eq.~\eqref{sources}.
\subsubsection{Black Hole}
For $m=m^+$, the metric function
\begin{equation}
f^+
=
-\left[
1-\left(\frac{r}{r_{\rm s}}\right)^n
\right]^k
\ ,
\end{equation}
with $f^+(r_{\rm s})=0$ and the metric signature is $(+-++)$ for $r<r_{\rm s}$ if $n>0$.
In fact, the density $\epsilon(r)$ will decrease for increasing
$r$ only if 
{\em i)}
$k=1$ and $n\in[0,1)$
or
{\rm ii)}
$k>1$ and $n\in(1,2]$.
Even though we should expect some energy conditions are
violated~\cite{Martin-Moruno:2017exc}, we find that the dominant energy condition,
\begin{equation}
{\epsilon}
\geq
0
\ ,
\quad
\epsilon
\geq
|{p}_{i}|
\quad
\left(i=r,\theta\right)
\ ,
\label{dom1}
\end{equation}
holds for $k>6$, whereas the strong energy condition
\begin{equation}
\epsilon+{p}_{r}+2\,{p}_{\theta}
\geq
0
\ ,
\quad
\epsilon+{p}_{i}
\geq
0
\quad
\left(i=r,\theta\right)
\ ,
\label{strong01}
\end{equation}
is satisfied for $k=1$.
\subsubsection{Mimicker}
For $m=m^-$, the metric function
\begin{equation}
f^-
=
-f^+
\ ,
\end{equation}
and the metric signature is $(-+++)$ for $r<r_{\rm s}$ if again $n>0$.
The density gradient and compactness are proportional to $k$,
with $k=1$ being the case for an isotropic object of uniform density
(incompressible fluid).
A monotonic decrease of the density $\epsilon(r)$ with increasing $r$
is only possible for  $k>1$ and $n\in(1,2]$.
The dominant energy condition is satisfied for $k>3$ and $n=2$.
\subsubsection{De~Sitter and anti-de~Sitter}
It is straightforward to interpret the geometry determined by the mass
function in Eq.~\eqref{m} for $n=2$ in terms of vacuum energy.
First of all, we notice that $k=1$ and $n=2$ yield the curvature scalar
\begin{equation}
\label{scalar1}
R
=
\left\{
\begin{array}{ll}
\strut\displaystyle\frac{4}{r^2}-4\,\Lambda
\quad
&
{\rm for}\ m=m^+ \ {\rm (BH)}
\\
\\
4\,\Lambda
&
{\rm for}\ m=m^- \ {\rm (mimicker)}
\ ,
\end{array}
\right.
\end{equation}
where
\begin{equation}
\Lambda=3/\rs^2
\label{Lrs}
\end{equation}
is the (effective) cosmological constant.
These expressions correspond to anti-de~Sitter (AdS) spacetime
filled with some matter producing a scalar singularity at the origin and
de~Sitter spacetime, respectively.
\par
For $n=2$ and any real $k>1$, we obtain deformations of the two basic
configurations in Eq.~\eqref{scalar1} and we can argue
that $k>1$ parameterises deviations from dS or AdS.
In particular, for $k\in\mathbb{N}$, the deformed dS or AdS have a simple
interpretation in terms of compositions of configurations of the kind
in Eq.~\eqref{gdks} with Eq.~\eqref{m}. 
This can be seen clearly from the energy density~\eqref{sources}
for the two cases $m^\pm$ in Eq.~\eqref{m}, namely
\begin{eqnarray}
\label{super}
\kappa\,\epsilon^+_k
&=&
\frac{2}{r^2}
+k!\,
\sum_{p=1}^{k}
\,(-1)^{p}\,
\frac{2\,p+1}{p!\,(k-p)!}\,
\frac{r^{2\,(p-1)}}{\rs^{2\,p}}
\nonumber
\\
&=&
\frac{1}{r^2}
-k\,\Lambda
+\epsilon^{(2)}_k
-\epsilon^{(3)}_k
+
\ldots
+(-1)^k\,\epsilon^{(k)}_k
\end{eqnarray}
and
\begin{eqnarray}
\label{super2}
\kappa\,\epsilon^-_k
&=&
k!\,
\sum_{p=1}^{k}
(-1)^{p+1}\,
\frac{2\,p+1}{p!\,(k-p)!}\,
\frac{r^{2\,(p-1)}}{\rs^{2\,p}}
\nonumber
\\
&=&
k\,\Lambda
-\epsilon^{(2)}_k
+\ldots
+(-1)^{k+1}\,\epsilon^{(k)}_k
\ ,
\end{eqnarray}
where 
\begin{equation}
\epsilon^{(1<p\le k)}_k
\equiv
\frac{k!\,(2\,p+1)}{p!\,(k-p)!\,r^2}
\left(\frac{r}{r_{\rm s}}\right)^{2\,p}
\end{equation}
and $\epsilon^{(1)}_k=\Lambda$, with $\Lambda$ defined in Eq.~\eqref{Lrs}.
Since the $\epsilon^{(p)}_k$ appear with alternating signs,
we can interpret~\eqref{super} as a superposition of ``fluctuations''
around the basic dS and AdS configurations.
\par
It is important to remark that the leading behaviour of $\epsilon_k^-$
for $r\to 0$ is always given by a constant dS-like term, whereas
$\epsilon_k^+\sim r^{-2}$.
The latter result confirms that the BH metric is integrable, so that
one indeed has
\begin{equation}
m^\pm(r)
=
4\,\pi\,\kappa
\int_0^r
\bar r^2\,d \bar r\,\epsilon_k^\pm(\bar r)
\ ,
\label{meps}
\end{equation}
both for mimickers and BHs (see Appendix~\ref{A:pg} for more details).
\subsection{Exterior}
\label{sec:exterior}
\begin{figure}[t!]
\centering
\includegraphics[width=0.45\textwidth]{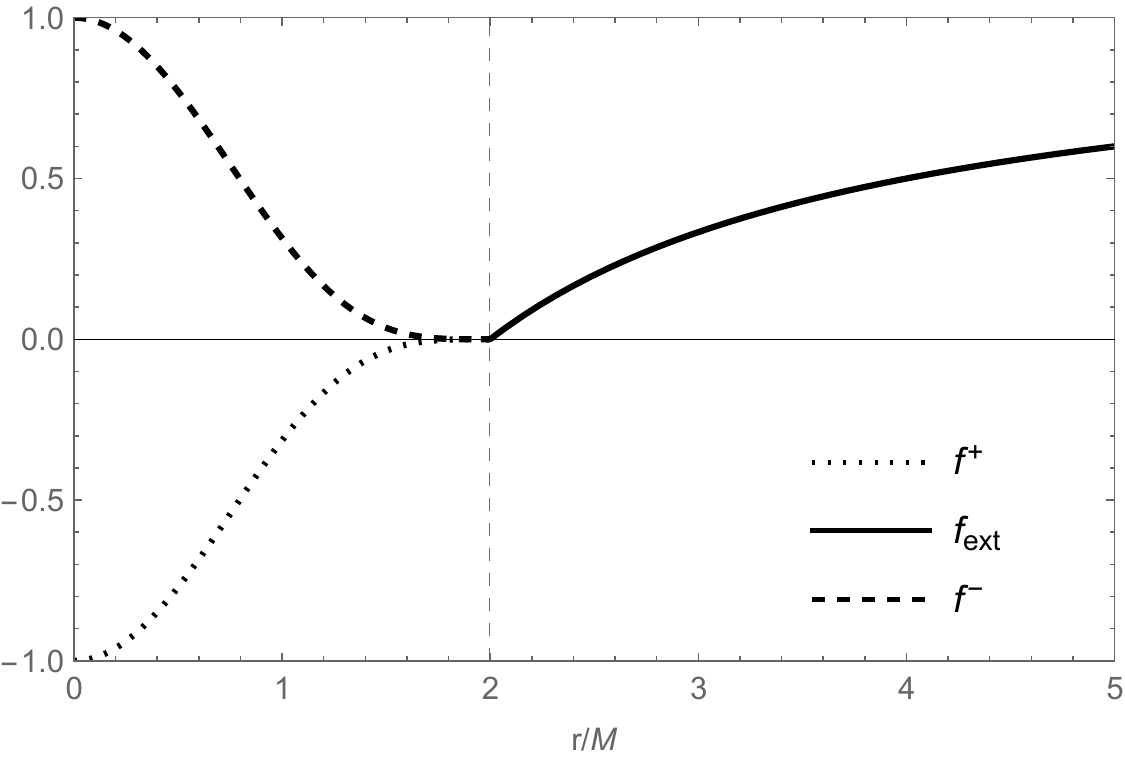}  
\caption{Metric function $f=f(r)$ for BH (dotted line) and mimicker (dashed line)
for ${\cal M}/M\approx\,1.0$, $n=2$ and $k=4$.
Dashed vertical line represents $r=\rs$ (in units of $\cal M$).}
\label{figfinext}
\end{figure}
\begin{figure}[t!]
\centering
\includegraphics[width=0.45\textwidth]{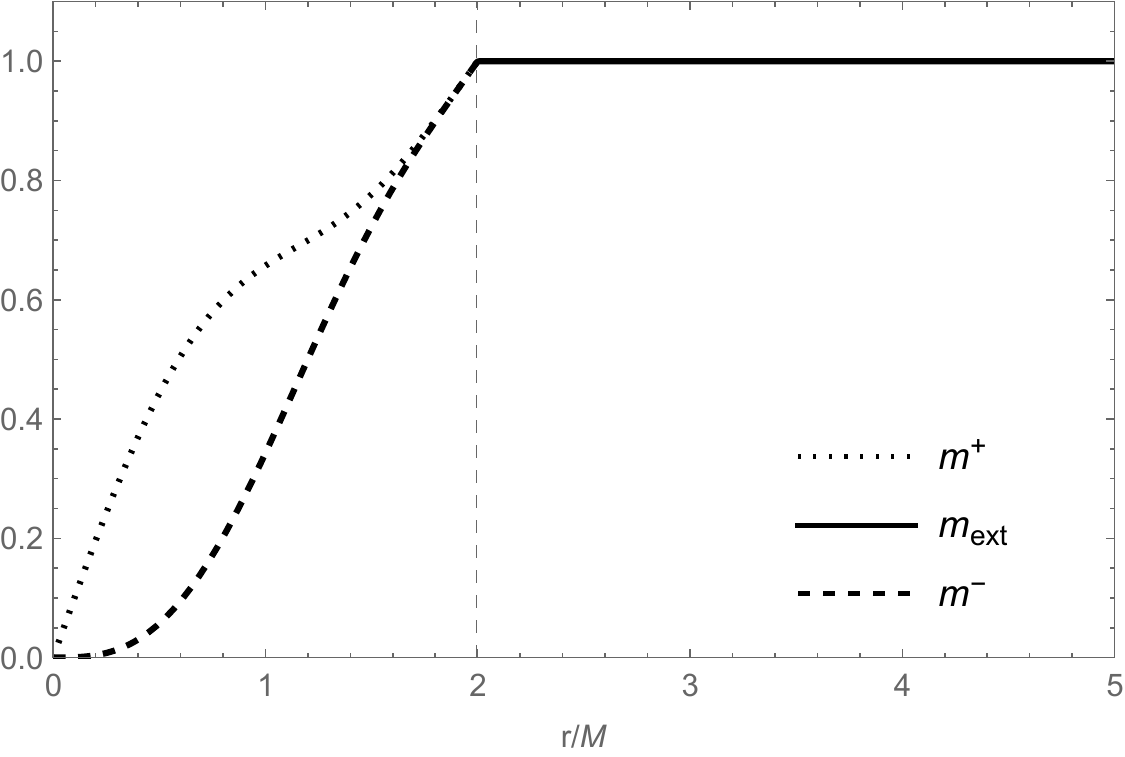}  
\caption{Mass function $m=m(r)$ for BH (dotted line) and mimicker (dashed line)
for ${\cal M}/M\approx\,1.0$, $n=2$ and $k=4$.
Dashed vertical line represents $r=\rs$ (in units of $\cal M$).}
\label{figminext}
\end{figure}
The next step is to extend our solutions~\eqref{metric} with $\Phi=0$
and $m=m^\pm(r)$ to the region $r>\rs$.
First of all, it is easy to prove that these solutions cannot be smoothly joined
with the Schwarzschild vacuum at $r=\rs$~\cite{Ovalle:2023vvu}.
In fact, the metric of this vacuum is also of the form in Eq.~\eqref{metric} with $\Phi=0$
and constant mass function $m=\mathcal M$.
Hence, $m'(\rs^+)=0$ cannot equal the derivative of $m^\pm(\rs^-)$.
\par
An exterior metric (for $r>\rs$) of the form in Eq.~\eqref{metric} with $\Phi=0$,
which smoothly matches the interiors with the mass functions $m=m^\pm(r)$ at $r=\rs$,
and approaches the Schwarzschild metric for $r\gtrsim\rs$,
is given by~\footnote{For the details, see Refs.~\cite{Ovalle:2023vvu,Ovalle:2020kpd}.}
\begin{equation}
f_{\rm ext}
=
1
-\frac{{2\,\cal M}}{r}
-\left(1-\frac{{2\,\cal M}}{\rs}\right)
\exp\left\{-2\,\frac{r-\rs}{2\,\mathcal{M}-\ell}\right\}
\ ,
\label{mimickerexterior}
\end{equation}
where $\mathcal{M}$ is the ADM mass (measured by an asymptotic observer)
and $\ell$ a length scale which must satisfy {${\cal M}\leq\ell<2\,{\cal M}$}
in order to ensure asymptotic flatness and also have
\begin{equation}
\label{horizon}
\rs
=
2\,M
=
{\cal M}+\sqrt{\ell\,{\cal M}-{\cal M}^2}
\ .
\end{equation}
\par
It is important to remark the difference between ${\cal M}$ and $M$
in Eq.~\eqref{Mmim}.
The former is the total mass of the configuration, while the latter is the fraction
of mass confined within the region $r\leq\rs$.
Indeed, from the expression in Eq.~\eqref{horizon} we see that
$M<{\cal M}\le 2\,M$, with
\begin{eqnarray}
\label{limit}
\ell\to 2\,{\cal M}
\quad
\Rightarrow
\quad
{\cal M}\to M
\ .
\end{eqnarray}
Hence we conclude that $\ell$ controls the amount of mass ${\cal M}$ contained
within the trapping surface $r=\rs$.
The case $\ell\to 2\,{\cal M}$ in Eq.~\eqref{limit} would correspond to the
Schwarzschild BH with the total mass ${\cal M}=M$ confined within the region
$r\leq\rs$.
However, we have seen that this limiting case is excluded and we can conclude
that both integrable BHs and mimickers are dressed with a shell of matter of
(arbitrarily small) thickness $\Delta\simeq 2\,{\cal M}-\ell>0$ and (arbitrarily small)
mass ${\cal M}-M\simeq\Delta/4$ (for example, $\Delta/r_{\rm s}\simeq 0.05$
in Figs.~\ref{figfinext} and~\ref{figminext}).
\subsection{Complete geometries}
By matching the interior for $m=m^+$ with the exterior~\eqref{mimickerexterior}
one obtains a complete BH geometry.
Likewise, a mimicker is obtained by joining the exterior~\eqref{mimickerexterior}
to the interior with $m=m^-$ (see Figs.~\ref{figfinext} and~\ref{figminext} for an 
example).
In particular, we note that the continuity of the mass function across $\rs=2\,M$
means that $f_{\rm int}^\pm(\rs^-)=0=f_{\rm ext}(\rs^+)$ and continuity of $m'$
implies that $f_{\rm ext}'(\rs^+)=0=(f_{\rm int}^\pm)'(\rs^-)$.
\par
We would like to highlight a few more features of these two solutions:
\begin{itemize}
\item
Both solutions contain only two charges, {\em i.e.}~the ADM mass
${\cal M}$ and the length scale $\ell=2\,\mathcal{M}-\Delta$.
\item
A BH and a mimicker can have the same exterior geometry with ADM mass
$\cal M$ provided the interior mass $M=m^\pm(\rs)$ (equivalently $\ell$)
is also the same.
\item
The sphere $r=\rs$ is an event horizon for the BH and an infinite redshift
hypersurface for the mimicker.
\item
The interior $r\leq\rs$ is quite different in the two cases:
the BH has one horizon (there is no Cauchy inner horizon),
and contains an integrable singularity at $r=0$ [see Eq.~\eqref{scalar1}];
the mimicker is completely regular inside.
\end{itemize}
We also remark that there is a discontinuity in the tangential pressure
$p_\theta\sim m''$ at $r=r_{\rm s}$~\cite{Ovalle:2023vvu,Ovalle:2020kpd}.
\section{From de~Sitter to anti-de~Sitter}
\label{sec4}
We have seen that the the mass functions in Eq.~\eqref{m} can correspond
to static BHs and mimickers.
We can now consider the possibility of dynamical processes that result
in the transition between two such configurations.
In general, the ADM mass $\cal M$ and radius $r_{\rm s}=2\,M$ (equivalently,
the length scale $\ell$) could change in time.
However, it is easier to consider cases in which both parameters remain
constant.
\par
\begin{figure}[t!]
	\centering
	\includegraphics[width=0.45\textwidth]{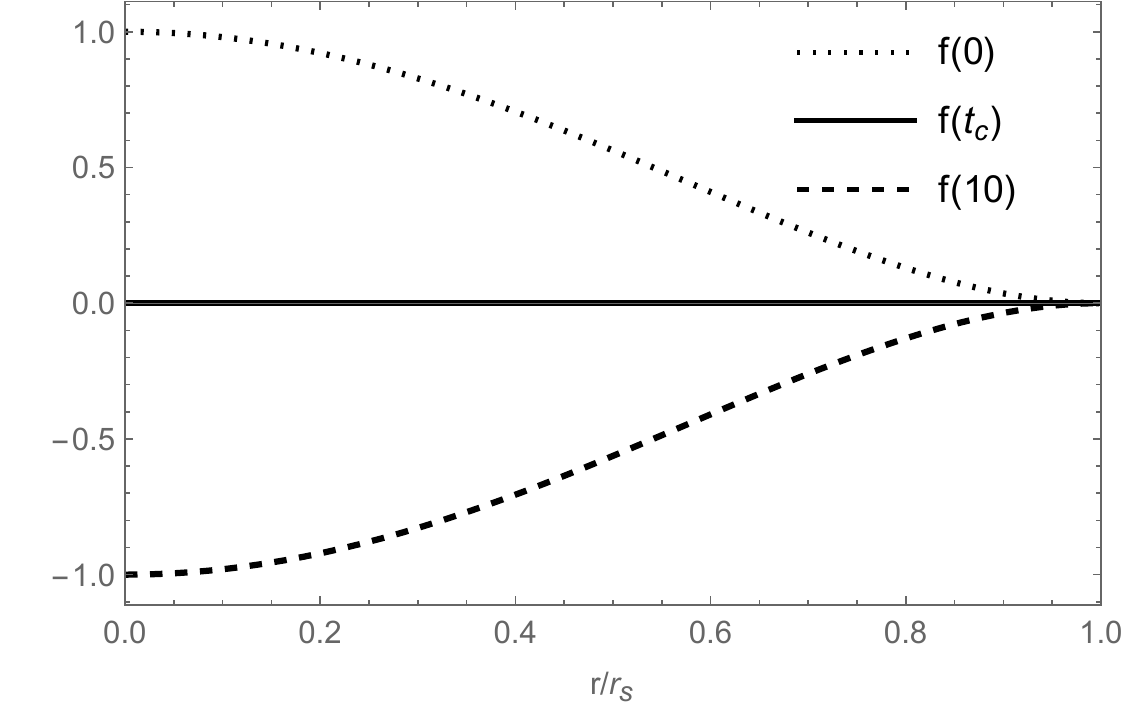}  
	\caption{Metric function $f$ for mimicker-to-BH with $k=2$ at different times
		($t$ in units of $\omega^{-1}$) and $t_{\rm c}$ in Eq.~\eqref{tc}.}
	\label{fig2}
\end{figure}
\begin{figure}[t!]
	\centering
	\includegraphics[width=0.45\textwidth]{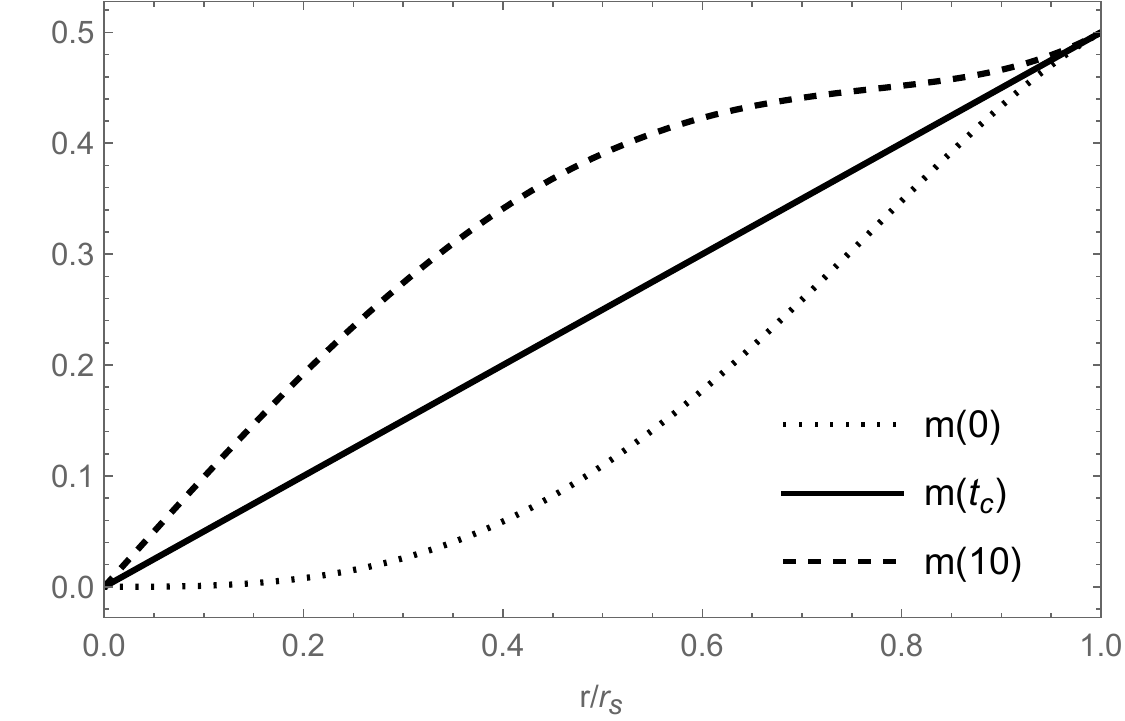}  
	\caption{Mass function $m$ for mimicker-to-BH with $k=2$ at different times
		($t$ in units of $\omega^{-1}$) and $t_{\rm c}$ in Eq.~\eqref{tc}.}
	\label{fig3}
\end{figure}
In particular, since the BH metric with mass function $m=m^+$
involves a larger fraction of the total mass near the centre
than the mimicker with $m=m^-$ (see Fig.~\ref{figminext}), it makes sense
to assume that the mimicker represents the initial configuration and the BH
is the final configuration for this process (the opposite may be more interesting
for cosmology, see Appendix~\ref{A:inside}).
This means that the mass function for $0\le r\leq\rs$ must be time dependent,
$m=m(r,t)$ (see also Appendix~\ref{A:time}),
and start from the mimicker with $f=f^-$, say at $t=0$,
\begin{equation}
m(r,t=0)
=
m^{-}(r;r_{\rm s})
\ .
\end{equation}
to evolve into the BH with $f=f^+$, at least in an infinite amount of time, 
\begin{equation}
m(r,t\to \infty)
=
m^{+}(r;r_{\rm s})
\ ,
\end{equation}
with $r_{\rm s}=2\,M$ at all times.
The exterior geometry is instead static and described by $f_{\rm ext}$
in Eq.~\eqref{mimickerexterior} with constant $\cal M$ and $\ell$ related
with $r_{\rm s}$ according to Eq.~\eqref{horizon}.
\par
An example of a time-dependent mass function for the interior with the
above features is given by
\begin{equation}
\label{m(t)}
m(r,t)
=
\frac{r}{2}
\left\{
1+\left(1-2\,e^{-\omega\,t}\right)
\left[1-\left(\frac{r}{\rs}\right)^2\right]^{k}
\right\}
\ ,
\end{equation}
where $\omega^{-1}$ is a time scale associated with the transition.
In this respect, it is interesting to note that the complete spacetime
metric changes signature across $r=\rs$ (and the surface $r=\rs$
becomes a horizon) at a time 
\begin{equation}
\label{tc}
t_{\rm c}=\ln(2)\,\omega^{-1}
\ ,
\end{equation}
when $f(r)=0$ and $m=r/2$ for $0\le r\le \rs$
(see Figs.~\ref{fig2} and \ref{fig3} for an example). 
\par
Since the exterior geometry does not change, one can look at this
process as being consistent with the fact that $r=\rs$ is a sphere of infinite
redshift for all $t>0$.
Mechanisms to allow for energy loss must therefore involve quantum
effects, like the Hawking evaporation~\cite{Hawking:1975vcx},
which we have neglected here.
\section{Conclusion and final remarks}
\label{conc}
Studying the possible evolution from (horizonless) ultracompact objects
to BHs using purely analytical and exact models, is a great challenge.
In this sense, the model represented by the mass function in Eq.~\eqref{m(t)}
could be pioneering in this scenario.
Its generalisation to more complex and realistic situations could help
to shed light on new aspects of the gravitational collapse, in particular,
on the existence of ultracompact stellar configurations as the final stage.
\par
Although Eq.~\eqref{m(t)} represents an advancement,
it is fair to mention that our model is not free from limitations. 
One of them is the fact that external observers could never detect
this specific transition from a mimicker to the BH in the classical theory.
This is a direct consequence of the two (Cauchy horizon free)
configurations in Eq.~\eqref{m}, which represent the initial and final states
of our model, respectively.
We can see that the BH horizon always coincides with the infinite redshift
surface of the mimicker and, correspondingly, no energy is emitted (classically)
during the process.
In this form, our model is still a valid starting point to investigate
near-horizon {\em quantum\/} physics.
\par
We would like to conclude by emphasizing that we do not mean
to provide a phenomenologically complete model.
On the contrary, our objective is, more humbly but no less importantly,
to lay the foundation for analytically exploring the collapse of
ultra-compact configurations into BHs.
In this perspective, there are many aspects that deserve further
studying, such as its stability and extension to include energy emission
and rotating systems.
One should also consider alternative transitions from our mimickers,
which are regular objects, to non-singular (rather than integrable) BHs,
which will generically contain two horizons, namely the event horizon
and the Cauchy horizon. All such aspects go beyond the scope of this article.
\subsection*{Acknowledgments}
J.O.~is partially supported by ANID FONDECYT Grant No.~1210041.
R.C.~and A.K.~are partially supported by the INFN grant FLAG.
The work of R.C.~has also been carried out in the framework of activities of the
National Group of Mathematical Physics (GNFM, INdAM).
\appendix
\section{Painlev\'e-Gullstrand coordinates}
\label{A:pg}
For the metric~\eqref{metric} with $\Phi=0$, one can introduce a
Painlev\'e-Gullstrand time $T$ such that spatial hypersurfaces of constant $T$
are flat,
\begin{equation}
d s^2
=
-
f\,d T^2
+
2\,\sqrt{1-f}\,d T\,d r
+
d r^2
+
r^2\,d\Omega^2
\ .
\label{gGP}
\end{equation}
We can next introduce tetrads which, for the angular part, read
\begin{eqnarray}
e_{(2)}^\mu
&=&
\left(
0,0,\frac{1}{r},0
\right)
\nonumber
\\
e_{(3)}^\mu
&=&
\left(
0,0,0,\frac{1}{r\,\sin\theta}
\right)
\ .
\end{eqnarray}
Where $0<f\le 1$, like inside the mimicker with $f=f^-$ and in the exterior with
$f=f_{\rm ext}$, one can define two tetrads
\begin{eqnarray}
e_{(0)}^\mu
&=&
\left(
\frac{1}{\sqrt{f}},0,0,0
\right)
\\
e_{(1)}^\mu
&=&
\left(
-{\sqrt{\frac{1}{f}-1}},\sqrt{f},0,0
\right)
\ .
\end{eqnarray}
Where $f<0$, like inside the BH with $f=f^+$, one can instead use 
\begin{eqnarray}
e_{(0)}^\mu
&=&
\left(
-\sqrt{1-\frac{1}{f}},\sqrt{-f},0,0
\right)
\\
e_{(1)}^\mu
&=&
\left(
\frac{1}{\sqrt{-f}},0,0,0
\right)
\ .
\end{eqnarray}
\par
The effective energy-momentum tensor sourcing the metric can then
be obtained by projecting the Einstein tensor on the tetrad.
From
\begin{eqnarray}
G^T_{\ T}
&=&
G^r_{\ r}
=
\frac{f+r\,f'-1}{r^2}
\nonumber
\\
G^\theta_{\ \theta}
&=&
G^\phi_{\ \phi}
=
\frac{2\,f'+r\,f''}{2\,r}
\ ,
\end{eqnarray}
one finds
\begin{equation}
\kappa\,\epsilon
=
{G_{\mu\nu}\,e^\mu_{(0)}\,e^\nu_{(0)}}
=
\frac{2\,m'}{r^2}
\ .
\end{equation}
Since the spatial metric is flat, it is now easy to see that the total energy
within a sphere of radius $r$ at constant $T$ is indeed given by Eq.~\eqref{meps},
regardless of the sign of $f$.
Moreover the spatial volume of these hypersurfaces inside $\rs$ is also constant
and equals $(4/3)\,\pi\,\rs^3$.
Furthermore, the radial pressure
\begin{equation}
\kappa\,p_r
=
{G_{\mu\nu}\,e^\mu_{(1)}\,e^\nu_{(1)}}
=
-\kappa\,\epsilon
\ ,
\end{equation}
and the tangential pressure
\begin{equation}
\kappa\,p_\theta
=
G_{\mu\nu}\,e_{(2)}^\mu\,e_{(2)}^\nu
=
G_{\mu\nu}\,e_{(3)}^\mu\,e_{(3)}^\nu
=
\kappa\,p_\phi
=
-\frac{m''}{r}
\ .
\end{equation}
All expressions are clearly in agreement with Eq.~\eqref{sources}.
\section{Inside the black hole}
\label{A:inside}
Let us consider the geometry inside the horizon $r=\rs$
for $n=2$ and $k=1$ [see Eq.~\eqref{scalar1}] as a whole universe,
in the spirit of Ref.~\cite{Doran:2006dq}.
In this case the metric reads
\begin{equation}
ds^2
=
\frac{dt^2}{1-{t^2}/{t_0^2}}
-
\left(1-\frac{t^2}{t_0^2}\right)dr^2
-
t^2\,d\Omega^2
\ .
\label{inside}
\end{equation}
The components of the Ricci tensor are
\begin{eqnarray}
R_{\ t}^t
&=&
R_{\ r}^r
=
\frac{3}{t_0^2}
\nonumber
\\
R_{\ \theta}^{\theta}
&=&
R_{\ \phi}^{\phi}
=
-\frac{2}{t^2}+\frac{3}{t_0^2}
\ ,
\label{Ricci}
\end{eqnarray}
and the Ricci scalar is
\begin{equation}
R
=
\frac{12}{t_0^2}-\frac{4}{t^2}
\ .
\label{Ricci1}
\end{equation}
The components of the Einstein tensor therefore read 
\begin{eqnarray}
G_{\ t}^t
&=&
G_{\ r}^r
=
-\frac{3}{t_0^2}+\frac{2}{t^2}
\nonumber
\\
G_{\ \theta}^{\theta}
&=&
G_{\ \phi}^{\phi}
=
-\frac{3}{t_0^2}
\ .
\label{Ricci2}
\end{eqnarray}
Comparing Eqs.~\eqref{Ricci2} with the expression~\eqref{emt} for the
energy-momentum tensor, we see that this universe is filled with a negative
cosmological constant and an anisotropic fluid, in agreement with the analysis
in Appendix~\ref{A:pg}.
\par
We can rewrite the metric~\eqref{inside} in terms of the cosmic time
$\tau$ with
\begin{equation}
t = t_0\,\sin\left(\frac{\tau}{t_0}\right)
\ ,
\label{time}
\end{equation}
which yields
\begin{equation}
ds^2
=
d\tau^2
-\cos^2\left(\frac{\tau}{t_0}\right)
dr^2
-t_0^2\,\sin^2\left(\frac{\tau}{t_0}\right)
d\Omega^2
\ .
\label{inside1}
\end{equation}
This metric describes a Kantowski-Sachs universe with a simple form
for the two scale factors~\cite{Kantowski:1966te,Brehme:1977fi}.
\section{Time-dependent energy-momentum tensor}
\label{A:time}
The components of the energy-momentum tensor for a metric of the form~\eqref{metric}
with $\Phi=0$ and $m=m(r,t)$ can be easily obtained from the Einstein equations~\eqref{efe}
and read
\begin{eqnarray}
\label{sourcesm(t)}
T^0_{\ 0}
&=&
\frac{2\,{m}'}{\kappa\,r^2}
\ ,
\quad
T^1_{\ 1}
=
-\frac{2\,{m}'}{\kappa\,r^2}
=
-T^0_{\ 0}
\ ,
\nonumber
\\
T^2_{\ 2}
&=&
-\frac{{m}''}{\kappa\,r}-\frac{4\,r\,{\dot m}^2}{\kappa\,(r-2\,m)^3}-\frac{r\,{\ddot m}^2}{\kappa\,(r-2\,m)^2}
\ ,
\end{eqnarray}
where dots denote derivatives with respect to $t$.
These expressions reduce to those in Appendix~\ref{A:pg} for the static case $\dot m=0$.
Moreover, one also finds a flux of energy 
\begin{equation}
T^0_{\ 1}
=
-\frac{2\,{\dot m}}{\kappa\,(r-2\,m)^2}
\ ,
\end{equation}
which does not appear in the static case. 
\par
The (apparently) singular behaviour of terms containing $\dot m$ and $\ddot m$
for $r\to 2\,m$ is just due to the choice of Schwarzschild-like coordinates
in Eq.~\eqref{metric}.
One can remove this apparent singularity by employing Eddington-Finkelstein
coordinates, in which the metric reads
\begin{equation}
\label{Eddington}
ds^{2}
=
-f\,dv^{2}
+2\,dv\,dr
+r^2\,d\Omega^2
\ ,
\end{equation}
and the components of the energy-momentum tensor equal those in
Eq.~\eqref{sources} with $T^0_{\ 1}=0$.
\bibliography{references.bib}

\begin{thebibliography}{23}%
\makeatletter
\providecommand \@ifxundefined [1]{%
 \@ifx{#1\undefined}
}%
\providecommand \@ifnum [1]{%
 \ifnum #1\expandafter \@firstoftwo
 \else \expandafter \@secondoftwo
 \fi
}%
\providecommand \@ifx [1]{%
 \ifx #1\expandafter \@firstoftwo
 \else \expandafter \@secondoftwo
 \fi
}%
\providecommand \natexlab [1]{#1}%
\providecommand \enquote  [1]{``#1''}%
\providecommand \bibnamefont  [1]{#1}%
\providecommand \bibfnamefont [1]{#1}%
\providecommand \citenamefont [1]{#1}%
\providecommand \href@noop [0]{\@secondoftwo}%
\providecommand \href [0]{\begingroup \@sanitize@url \@href}%
\providecommand \@href[1]{\@@startlink{#1}\@@href}%
\providecommand \@@href[1]{\endgroup#1\@@endlink}%
\providecommand \@sanitize@url [0]{\catcode `\\12\catcode `\$12\catcode
  `\&12\catcode `\#12\catcode `\^12\catcode `\_12\catcode `\%12\relax}%
\providecommand \@@startlink[1]{}%
\providecommand \@@endlink[0]{}%
\providecommand \url  [0]{\begingroup\@sanitize@url \@url }%
\providecommand \@url [1]{\endgroup\@href {#1}{\urlprefix }}%
\providecommand \urlprefix  [0]{URL }%
\providecommand \Eprint [0]{\href }%
\providecommand \doibase [0]{http://dx.doi.org/}%
\providecommand \selectlanguage [0]{\@gobble}%
\providecommand \bibinfo  [0]{\@secondoftwo}%
\providecommand \bibfield  [0]{\@secondoftwo}%
\providecommand \translation [1]{[#1]}%
\providecommand \BibitemOpen [0]{}%
\providecommand \bibitemStop [0]{}%
\providecommand \bibitemNoStop [0]{.\EOS\space}%
\providecommand \EOS [0]{\spacefactor3000\relax}%
\providecommand \BibitemShut  [1]{\csname bibitem#1\endcsname}%
\let\auto@bib@innerbib\@empty
\bibitem [{\citenamefont {Mazur}\ and\ \citenamefont
  {Mottola}(2004)}]{Mazur:2004fk}%
  \BibitemOpen
  \bibfield  {author} {\bibinfo {author} {\bibfnamefont {P.~O.}\ \bibnamefont
  {Mazur}}\ and\ \bibinfo {author} {\bibfnamefont {E.}~\bibnamefont
  {Mottola}},\ }\href {\doibase 10.1073/pnas.0402717101} {\bibfield  {journal}
  {\bibinfo  {journal} {Proc. Nat. Acad. Sci.}\ }\textbf {\bibinfo {volume}
  {101}},\ \bibinfo {pages} {9545} (\bibinfo {year} {2004})},\ \Eprint
  {http://arxiv.org/abs/gr-qc/0407075} {arXiv:gr-qc/0407075} \BibitemShut
  {NoStop}%
\bibitem [{\citenamefont {Mazur}\ and\ \citenamefont
  {Mottola}(2023)}]{Mazur:2001fv}%
  \BibitemOpen
  \bibfield  {author} {\bibinfo {author} {\bibfnamefont {P.~O.}\ \bibnamefont
  {Mazur}}\ and\ \bibinfo {author} {\bibfnamefont {E.}~\bibnamefont
  {Mottola}},\ }\href {\doibase 10.3390/universe9020088} {\bibfield  {journal}
  {\bibinfo  {journal} {Universe}\ }\textbf {\bibinfo {volume} {9}},\ \bibinfo
  {pages} {88} (\bibinfo {year} {2023})},\ \Eprint
  {http://arxiv.org/abs/gr-qc/0109035} {arXiv:gr-qc/0109035} \BibitemShut
  {NoStop}%
\bibitem [{\citenamefont {Visser}\ and\ \citenamefont
  {Wiltshire}(2004)}]{Visser:2003ge}%
  \BibitemOpen
  \bibfield  {author} {\bibinfo {author} {\bibfnamefont {M.}~\bibnamefont
  {Visser}}\ and\ \bibinfo {author} {\bibfnamefont {D.~L.}\ \bibnamefont
  {Wiltshire}},\ }\href {\doibase 10.1088/0264-9381/21/4/027} {\bibfield
  {journal} {\bibinfo  {journal} {Class. Quant. Grav.}\ }\textbf {\bibinfo
  {volume} {21}},\ \bibinfo {pages} {1135} (\bibinfo {year} {2004})},\ \Eprint
  {http://arxiv.org/abs/gr-qc/0310107} {arXiv:gr-qc/0310107 [gr-qc]}
  \BibitemShut {NoStop}%
\bibitem [{\citenamefont {Lobo}(2006)}]{Lobo:2005uf}%
  \BibitemOpen
  \bibfield  {author} {\bibinfo {author} {\bibfnamefont {F.~S.~N.}\
  \bibnamefont {Lobo}},\ }\href {\doibase 10.1088/0264-9381/23/5/006}
  {\bibfield  {journal} {\bibinfo  {journal} {Class. Quant. Grav.}\ }\textbf
  {\bibinfo {volume} {23}},\ \bibinfo {pages} {1525} (\bibinfo {year}
  {2006})},\ \Eprint {http://arxiv.org/abs/gr-qc/0508115} {arXiv:gr-qc/0508115}
  \BibitemShut {NoStop}%
\bibitem [{\citenamefont {Lukash}\ and\ \citenamefont
  {Strokov}(2013)}]{Lukash:2013ts}%
  \BibitemOpen
  \bibfield  {author} {\bibinfo {author} {\bibfnamefont {V.~N.}\ \bibnamefont
  {Lukash}}\ and\ \bibinfo {author} {\bibfnamefont {V.~N.}\ \bibnamefont
  {Strokov}},\ }\href {\doibase 10.1142/S0217751X13500073} {\bibfield
  {journal} {\bibinfo  {journal} {Int. J. Mod. Phys. A}\ }\textbf {\bibinfo
  {volume} {28}},\ \bibinfo {pages} {1350007} (\bibinfo {year} {2013})},\
  \Eprint {http://arxiv.org/abs/1301.5544} {arXiv:1301.5544 [gr-qc]}
  \BibitemShut {NoStop}%
\bibitem [{\citenamefont {Casadio}(2022)}]{Casadio:2021eio}%
  \BibitemOpen
  \bibfield  {author} {\bibinfo {author} {\bibfnamefont {R.}~\bibnamefont
  {Casadio}},\ }\href {\doibase 10.1142/S0218271822501280} {\bibfield
  {journal} {\bibinfo  {journal} {Int. J. Mod. Phys. D}\ }\textbf {\bibinfo
  {volume} {31}},\ \bibinfo {pages} {2250128} (\bibinfo {year} {2022})},\
  \Eprint {http://arxiv.org/abs/2103.00183} {arXiv:2103.00183 [gr-qc]}
  \BibitemShut {NoStop}%
\bibitem [{\citenamefont {Casadio}\ \emph {et~al.}(2022)\citenamefont
  {Casadio}, \citenamefont {Giusti},\ and\ \citenamefont
  {Ovalle}}]{Casadio:2022ndh}%
  \BibitemOpen
  \bibfield  {author} {\bibinfo {author} {\bibfnamefont {R.}~\bibnamefont
  {Casadio}}, \bibinfo {author} {\bibfnamefont {A.}~\bibnamefont {Giusti}}, \
  and\ \bibinfo {author} {\bibfnamefont {J.}~\bibnamefont {Ovalle}},\ }\href
  {\doibase 10.1103/PhysRevD.105.124026} {\bibfield  {journal} {\bibinfo
  {journal} {Phys. Rev. D}\ }\textbf {\bibinfo {volume} {105}},\ \bibinfo
  {pages} {124026} (\bibinfo {year} {2022})},\ \Eprint
  {http://arxiv.org/abs/2203.03252} {arXiv:2203.03252 [gr-qc]} \BibitemShut
  {NoStop}%
\bibitem [{\citenamefont {Casadio}\ \emph {et~al.}(2023)\citenamefont
  {Casadio}, \citenamefont {Giusti},\ and\ \citenamefont
  {Ovalle}}]{Casadio:2023iqt}%
  \BibitemOpen
  \bibfield  {author} {\bibinfo {author} {\bibfnamefont {R.}~\bibnamefont
  {Casadio}}, \bibinfo {author} {\bibfnamefont {A.}~\bibnamefont {Giusti}}, \
  and\ \bibinfo {author} {\bibfnamefont {J.}~\bibnamefont {Ovalle}},\ }\href
  {\doibase 10.1007/JHEP05(2023)118} {\bibfield  {journal} {\bibinfo  {journal}
  {JHEP}\ }\textbf {\bibinfo {volume} {05}},\ \bibinfo {pages} {118} (\bibinfo
  {year} {2023})},\ \Eprint {http://arxiv.org/abs/2303.02713} {arXiv:2303.02713
  [gr-qc]} \BibitemShut {NoStop}%
\bibitem [{\citenamefont {Carballo-Rubio}\ \emph {et~al.}(2023)\citenamefont
  {Carballo-Rubio}, \citenamefont {Di~Filippo}, \citenamefont {Liberati},\ and\
  \citenamefont {Visser}}]{Carballo-Rubio:2023mvr}%
  \BibitemOpen
  \bibfield  {author} {\bibinfo {author} {\bibfnamefont {R.}~\bibnamefont
  {Carballo-Rubio}}, \bibinfo {author} {\bibfnamefont {F.}~\bibnamefont
  {Di~Filippo}}, \bibinfo {author} {\bibfnamefont {S.}~\bibnamefont
  {Liberati}}, \ and\ \bibinfo {author} {\bibfnamefont {M.}~\bibnamefont
  {Visser}},\ }\href@noop {} {\  (\bibinfo {year} {2023})},\ \Eprint
  {http://arxiv.org/abs/2302.00028} {arXiv:2302.00028 [gr-qc]} \BibitemShut
  {NoStop}%
\bibitem [{\citenamefont {Visser}(1995)}]{Visser:1995cc}%
  \BibitemOpen
  \bibfield  {author} {\bibinfo {author} {\bibfnamefont {M.}~\bibnamefont
  {Visser}},\ }\href@noop {} {\emph {\bibinfo {title} {{Lorentzian wormholes:
  From Einstein to Hawking}}}}\ (\bibinfo {year} {1995})\BibitemShut {NoStop}%
\bibitem [{\citenamefont {Misner}\ and\ \citenamefont
  {Sharp}(1964)}]{Misner:1964je}%
  \BibitemOpen
  \bibfield  {author} {\bibinfo {author} {\bibfnamefont {C.~W.}\ \bibnamefont
  {Misner}}\ and\ \bibinfo {author} {\bibfnamefont {D.~H.}\ \bibnamefont
  {Sharp}},\ }\href {\doibase 10.1103/PhysRev.136.B571} {\bibfield  {journal}
  {\bibinfo  {journal} {Phys. Rev.}\ }\textbf {\bibinfo {volume} {136}},\
  \bibinfo {pages} {B571} (\bibinfo {year} {1964})}\BibitemShut {NoStop}%
\bibitem [{\citenamefont {Hernandez}\ and\ \citenamefont
  {Misner}(1966)}]{Hernandez:1966zia}%
  \BibitemOpen
  \bibfield  {author} {\bibinfo {author} {\bibfnamefont {W.~C.}\ \bibnamefont
  {Hernandez}}\ and\ \bibinfo {author} {\bibfnamefont {C.~W.}\ \bibnamefont
  {Misner}},\ }\href {\doibase 10.1086/148525} {\bibfield  {journal} {\bibinfo
  {journal} {Astrophys. J.}\ }\textbf {\bibinfo {volume} {143}},\ \bibinfo
  {pages} {452} (\bibinfo {year} {1966})}\BibitemShut {NoStop}%
\bibitem [{\citenamefont {Kerr}\ and\ \citenamefont
  {Schild}(1965)}]{kerrchild}%
  \BibitemOpen
  \bibfield  {author} {\bibinfo {author} {\bibfnamefont {R.~P.}\ \bibnamefont
  {Kerr}}\ and\ \bibinfo {author} {\bibfnamefont {A.}~\bibnamefont {Schild}},\
  }\href {\doibase 10.1007/s10714-009-0857-z} {\bibfield  {journal} {\bibinfo
  {journal} {Proc. Symp. Appl. Math}\ }\textbf {\bibinfo {volume} {17}},\
  \bibinfo {pages} {199} (\bibinfo {year} {1965})}\BibitemShut {NoStop}%
\bibitem [{\citenamefont {Ovalle}(2017)}]{Ovalle:2017fgl}%
  \BibitemOpen
  \bibfield  {author} {\bibinfo {author} {\bibfnamefont {J.}~\bibnamefont
  {Ovalle}},\ }\href {\doibase 10.1103/PhysRevD.95.104019} {\bibfield
  {journal} {\bibinfo  {journal} {Phys. Rev.}\ }\textbf {\bibinfo {volume}
  {D95}},\ \bibinfo {pages} {104019} (\bibinfo {year} {2017})},\ \Eprint
  {http://arxiv.org/abs/1704.05899} {arXiv:1704.05899 [gr-qc]} \BibitemShut
  {NoStop}%
\bibitem [{\citenamefont {Ovalle}(2019)}]{Ovalle:2019qyi}%
  \BibitemOpen
  \bibfield  {author} {\bibinfo {author} {\bibfnamefont {J.}~\bibnamefont
  {Ovalle}},\ }\href {\doibase 10.1016/j.physletb.2018.11.029} {\bibfield
  {journal} {\bibinfo  {journal} {Phys. Lett.}\ }\textbf {\bibinfo {volume}
  {B788}},\ \bibinfo {pages} {213} (\bibinfo {year} {2019})},\ \Eprint
  {http://arxiv.org/abs/1812.03000} {arXiv:1812.03000 [gr-qc]} \BibitemShut
  {NoStop}%
\bibitem [{\citenamefont {Ovalle}(2023)}]{Ovalle:2023vvu}%
  \BibitemOpen
  \bibfield  {author} {\bibinfo {author} {\bibfnamefont {J.}~\bibnamefont
  {Ovalle}},\ }\href {\doibase 10.1103/PhysRevD.107.104005} {\bibfield
  {journal} {\bibinfo  {journal} {Phys. Rev. D}\ }\textbf {\bibinfo {volume}
  {107}},\ \bibinfo {pages} {104005} (\bibinfo {year} {2023})},\ \Eprint
  {http://arxiv.org/abs/2305.00030} {arXiv:2305.00030 [gr-qc]} \BibitemShut
  {NoStop}%
\bibitem [{\citenamefont {Arnowitt}\ \emph {et~al.}(1959)\citenamefont
  {Arnowitt}, \citenamefont {Deser},\ and\ \citenamefont
  {Misner}}]{Arnowitt:1959ah}%
  \BibitemOpen
  \bibfield  {author} {\bibinfo {author} {\bibfnamefont {R.~L.}\ \bibnamefont
  {Arnowitt}}, \bibinfo {author} {\bibfnamefont {S.}~\bibnamefont {Deser}}, \
  and\ \bibinfo {author} {\bibfnamefont {C.~W.}\ \bibnamefont {Misner}},\
  }\href {\doibase 10.1103/PhysRev.116.1322} {\bibfield  {journal} {\bibinfo
  {journal} {Phys. Rev.}\ }\textbf {\bibinfo {volume} {116}},\ \bibinfo {pages}
  {1322} (\bibinfo {year} {1959})}\BibitemShut {NoStop}%
\bibitem [{\citenamefont {Martin-Moruno}\ and\ \citenamefont
  {Visser}(2017)}]{Martin-Moruno:2017exc}%
  \BibitemOpen
  \bibfield  {author} {\bibinfo {author} {\bibfnamefont {P.}~\bibnamefont
  {Martin-Moruno}}\ and\ \bibinfo {author} {\bibfnamefont {M.}~\bibnamefont
  {Visser}},\ }\href {\doibase 10.1007/978-3-319-55182-1_9} {\bibfield
  {journal} {\bibinfo  {journal} {Fundam. Theor. Phys.}\ }\textbf {\bibinfo
  {volume} {189}},\ \bibinfo {pages} {193} (\bibinfo {year} {2017})},\ \Eprint
  {http://arxiv.org/abs/1702.05915} {arXiv:1702.05915 [gr-qc]} \BibitemShut
  {NoStop}%
\bibitem [{\citenamefont {Ovalle}\ \emph {et~al.}(2021)\citenamefont {Ovalle},
  \citenamefont {Casadio}, \citenamefont {Contreras},\ and\ \citenamefont
  {Sotomayor}}]{Ovalle:2020kpd}%
  \BibitemOpen
  \bibfield  {author} {\bibinfo {author} {\bibfnamefont {J.}~\bibnamefont
  {Ovalle}}, \bibinfo {author} {\bibfnamefont {R.}~\bibnamefont {Casadio}},
  \bibinfo {author} {\bibfnamefont {E.}~\bibnamefont {Contreras}}, \ and\
  \bibinfo {author} {\bibfnamefont {A.}~\bibnamefont {Sotomayor}},\ }\href
  {\doibase 10.1016/j.dark.2020.100744} {\bibfield  {journal} {\bibinfo
  {journal} {Phys. Dark Univ.}\ }\textbf {\bibinfo {volume} {31}},\ \bibinfo
  {pages} {100744} (\bibinfo {year} {2021})},\ \Eprint
  {http://arxiv.org/abs/2006.06735} {arXiv:2006.06735 [gr-qc]} \BibitemShut
  {NoStop}%
\bibitem [{\citenamefont {Hawking}(1975)}]{Hawking:1975vcx}%
  \BibitemOpen
  \bibfield  {author} {\bibinfo {author} {\bibfnamefont {S.~W.}\ \bibnamefont
  {Hawking}},\ }\href {\doibase 10.1007/BF02345020} {\bibfield  {journal}
  {\bibinfo  {journal} {Commun. Math. Phys.}\ }\textbf {\bibinfo {volume}
  {43}},\ \bibinfo {pages} {199} (\bibinfo {year} {1975})},\ \bibinfo {note}
  {[Erratum: Commun.Math.Phys. 46, 206 (1976)]}\BibitemShut {NoStop}%
\bibitem [{\citenamefont {Doran}\ \emph {et~al.}(2008)\citenamefont {Doran},
  \citenamefont {Lobo},\ and\ \citenamefont {Crawford}}]{Doran:2006dq}%
  \BibitemOpen
  \bibfield  {author} {\bibinfo {author} {\bibfnamefont {R.}~\bibnamefont
  {Doran}}, \bibinfo {author} {\bibfnamefont {F.~S.~N.}\ \bibnamefont {Lobo}},
  \ and\ \bibinfo {author} {\bibfnamefont {P.}~\bibnamefont {Crawford}},\
  }\href {\doibase 10.1007/s10701-007-9197-6} {\bibfield  {journal} {\bibinfo
  {journal} {Found. Phys.}\ }\textbf {\bibinfo {volume} {38}},\ \bibinfo
  {pages} {160} (\bibinfo {year} {2008})},\ \Eprint
  {http://arxiv.org/abs/gr-qc/0609042} {arXiv:gr-qc/0609042} \BibitemShut
  {NoStop}%
\bibitem [{\citenamefont {Kantowski}\ and\ \citenamefont
  {Sachs}(1966)}]{Kantowski:1966te}%
  \BibitemOpen
  \bibfield  {author} {\bibinfo {author} {\bibfnamefont {R.}~\bibnamefont
  {Kantowski}}\ and\ \bibinfo {author} {\bibfnamefont {R.~K.}\ \bibnamefont
  {Sachs}},\ }\href {\doibase 10.1063/1.1704952} {\bibfield  {journal}
  {\bibinfo  {journal} {J. Math. Phys.}\ }\textbf {\bibinfo {volume} {7}},\
  \bibinfo {pages} {443} (\bibinfo {year} {1966})}\BibitemShut {NoStop}%
\bibitem [{\citenamefont {Brehme}(1977)}]{Brehme:1977fi}%
  \BibitemOpen
  \bibfield  {author} {\bibinfo {author} {\bibfnamefont {R.~W.}\ \bibnamefont
  {Brehme}},\ }\href {\doibase 10.1119/1.10829} {\bibfield  {journal} {\bibinfo
   {journal} {Am. J. Phys.}\ }\textbf {\bibinfo {volume} {45}},\ \bibinfo
  {pages} {423} (\bibinfo {year} {1977})}\BibitemShut {NoStop}%
\end{thebibliography}%
\bibliographystyle{apsrev4-1.bst}
%
%
\end{document}